# Recommending Research Papers to Chemists: A Specialized Interface for Chemical Entity Exploration

Corinna Breitinger
University of Konstanz
& University of Göttingen
Germany
breitinger@gipplab.org

Kay Herklotz
University of Wuppertal
Germany
herklotz@uni-wuppertal.de

Tim Flegelskamp
University of Wuppertal
Germany
tim.flegelskamp@uni-wuppertal.de

Norman Meuschke
University of Göttingen
Germany
meuschke@gipplab.org

## ABSTRACT

Researchers and scientists increasingly rely on specialized information retrieval (IR) or recommendation systems (RS) to support them in their daily research tasks. Paper recommender systems are one such tool scientists use to stay on top of the ever-increasing number of academic publications in their field. Improving research paper recommender systems is an active research field. However, less research has focused on how the interfaces of research paper recommender systems can be tailored to suit the needs of different research domains. For example, in the field of biomedicine and chemistry, researchers are not only interested in textual relevance but may also want to discover or compare the contained chemical entity information found in a paper's full text. Existing recommender systems for academic literature do not support the discovery of this non-textual, but semantically valuable, chemical entity data. We present the first implementation of a specialized chemistry paper recommender system capable of visualizing the contained chemical structures, chemical formulae, and synonyms for chemical compounds within the document's full text. We review existing tools and related research in this field before describing the implementation of our ChemVis system. With the help of chemists, we are expanding the functionality of ChemVis, and will perform an evaluation of recommendation performance and usability in future work.

## CCS CONCEPTS

- Human-centered computing → Graphical user interfaces;
- Information systems → Recommender systems

## 1 Introduction

Scientists use search engines and recommender systems to quickly find the most relevant scientific literature in their field. Widely used paper recommender systems include the recommendations built-in to digital libraries like PubMed, IEEE, or ACM, as well as the recommendations integrated into academic search engines like Semantic Scholar and Google Scholar. However, dozens of prototypical recommender systems and hundreds of recommendation approaches have been proposed in the literature [1, 2].

Despite the strong research interest in improving academic paper recommendation, the interfaces of paper recommender systems have not been tailored to the specific needs of different research domains. This is surprising, when we consider that a) the properties of research papers and b) the information needs of scientists differ across disciplines.

For example, in the STEM fields: mathematical papers contain mathematical formulae as a form of text-independent feature [3], while biomedical and chemistry papers contain chemical formulae in a variety of formats, including pictorial representations of molecules or compounds, called structural diagrams (Figure 1).

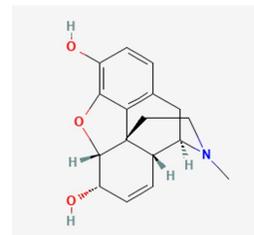

Figure 1: Structural diagram of morphine

In a biomedical or chemistry publication the following notations can be used to refer to the same compound: $C_{17}H_{19}NO_3$, C17H19NO3, and (5R,6S,9R,13S,14R)-4,5-Epoxy-N-methyl-7-morphinen-3,6-diol. The compound in question is *morphine*. To complicate matters further, authors may also substitute the compound's name with drug names, like MS Contin, Roxano, Oramorph, or synonyms like Morphium[1].

For chemists and biomedical researchers, it is time-consuming and challenging to manually find, compare, disambiguate, or look up the properties of chemical entities mentioned in a paper. To understand the challenge posed, we encourage readers to inspect the text of this chemistry publication [4]. This preliminary work paper proposes the first research paper recommender system tailored to the chemistry domain by supporting *chemical formula discovery and disambiguation* as a feature integrated into an interactive recommendation exploration interface.

---

[1] See: https://pubchem.ncbi.nlm.nih.gov/compound/Morphine

The contribution of our research is twofold. First, developing a paper recommendation prototype that extracts the chemical entities mentioned in the full text of research papers and displays their properties. Second, visualizing the chemical entities in a meaningful way in a graphical user interface.

## 2   Background and Related Work

Chemical entities can be described in text form, as chemical formulae, structure diagrams, or in chemical reaction schemes. For each of these formats, we examine exiting tools to automate the extraction of this data. ChemDataExtractor 2.0 supports the extraction of chemical entity names from the full text [5]. ChemSchematicResolver supports the extraction of chemical schematics [6], and ReactionDataExtractor extracts the chemical reactions from images [7]. We will briefly describe these toolkits and packages since they are a core component of our described prototype. They may also be of interest to researchers who work with digital libraries containing documents from the biomedical or chemistry domains. Finally, we will touch on related research before describing the implementation of our solution.

The ChemDataExtractor v2 [5] (CDE) [2] toolkit was introduced by the Molecular Engineering Group at the University of Cambridge to extract chemical information from research publications, theses, and patents. CDE provides a full pipeline for extracting chemical information from both text and tables in a domain-independent manner. Its capabilities include detecting chemical named entities, parsing of text and tables to identify chemical relationships, and resolving interdependencies between different elements.

ChemSchematicResolver[3] [6] is an open-source python package to extract chemical structures from schematic diagrams, as the one shown in Figure 1. This package uses an extension of CDE[4] for the automatic identification of candidate schematic diagrams. The input is a document in HTML or XML format, and the result is a machine-readable format of the schematic diagram's content, including labels. To achieve this, ChemSchematicResolver performs: image mining, feature detection, OCR, and resolving of the extracted structures.

Chemical reaction schemes encode particularly valuable semantic information since they can summarize the main contributions of a research paper or patent. For this extraction task, ReactionDataExtractor[5] [7] can be used to interpret chemical reaction schemes from images and convert them to a simplified molecular line-entry system (SMILES) string, which describes the chemical structure. ReactionDataExtractor uses DBSCAN in combination with statistical approaches and rule-based routines[6]. Finally, ImageDataExtractor[7] concludes the suite of chemistry-domain relevant packages. It automatically extracts quantitative data found in microscopy images, such as scale bars, occurrence of particles, and particle size.

To the best of our knowledge, no specialized paper recommender systems for the domain of chemistry exist. However, several related recommendation tasks are worth mentioning. For example, Savage et al. [8] proposed a system to help chemists in the task of identifying suitable candidate molecules (reactants) needed to synthesize a given target molecule (product). Recommender systems have also been proposed to make chemists in large pharmaceutical companies aware of each other's work and thus accelerate the drug discovery process [9]. Rohall et al. [10] match the drug crystal structures in the ongoing experiments of chemists with the structures of similar molecules in other experiments performed at the company. They describe five implementations of recommender systems to support chemists in different stages of their research and development workflow.

While no paper recommenders account for chemical entity information, many searchable molecular entity databases exist. *PubChem* [11] by the National Library of Medicine is the largest open-source database for retrieving chemical information. *ChemSpider* [12] by the Royal Society of Chemisty, or *ChEMBL* [13] and *ChEBI* [14] by the European Bioninformations Institute (EBI) are other freely available databases of molecular entities. However, there has been no integration of this knowledge into literature recommender systems.

## 3   ChemVis System

Our proposed ChemVis system for chemical formula exploration consists of a backend and frontend, each of which contains multiple components, as shown in Figure 2. The ChemVis backend is written in python. The frontend uses JavaScript and D3.js.

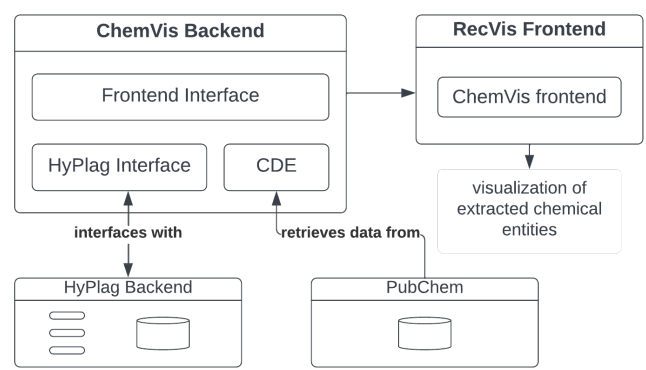

Figure 2: ChemVis System architecture

**Backend**

The ChemVis backend performs the work of extracting the chemical entities and their different representations from research papers using the CDE toolkit [5]. Our ChemVis backend communicates with the HyPlag backend via an HTTP

---

[2] http://www.chemdataextractor2.org/
[3] https://www.chemschematicresolver.org/
[4] https://github.com/edbeard/chemdataextractor-csr

[5] http://www.reactiondataextractor.org/
[6] https://github.com/dmw51/ReactionDataExtractor
[7] http://www.imagedataextractor.org/

API as shown in Figure 2. For a detailed description of our HyPlag system, please refer to [15]. HyPlag was developed to compute document similarity with the aim of detecting academic plagiarism. However, a range of HyPlag's functionality, such as PDF to XML file conversion, or the indexing, storage and retrieval of academic documents using an Elasticsearch Server, are identical to the functionality demanded by a paper recommender system such as *ChemVis*. Furthermore, the measures of global document similarity implemented in HyPlag are suitable for computing similarity for the paper recommendation use case. For these reasons, we have relied on HyPlag's existing architecture to fulfill these needs.

The upload of PDF files is performed with an HTTP POST request, which uploads the file to the HyPlag backend. The HyPlag backend returns a document ID, which is used to complete an HTTP GET request to get the XML file, which is then used to extract the different chemical entities of the scientific paper. The XML file output from the HyPlag backend is used for CDE. In this XML file, all different paper elements are tagged using XPath expressions. Since CDE uses a document class, which separates scientific papers into multiple elements like the titles, abstract, figures etc., the XML tags are used to identify these different elements of the document. After the matching of elements, the whole XML file is recursively parsed through all its elements.

As mentioned in the Background, PubChem is the largest open-source database for chemical information. It contains 111M compounds and 279M substances as of 2022. After the extraction of chemical entities from a scientific paper, we make use of this data to retrieve and display a specified set of properties for each chemical entity. We use two search modes. The first mode is using the name of the chemical and the second mode is using the chemical formula. We extract the following properties: IUPAC name of the chemical, PubChem ID, molecular structure, molecular formula, molecular weight. Other properties can be added quickly by adding the property's name to the list of extracted properties. The molecular structure is a PNG image encoded as a Base64 string. This string is then used by the frontend to display the image of the molecular structure of the chemical entity.

**Frontend**

The chemical entities extracted from the full text of the recommended papers are displayed in a graphical user interface. Figure 3 shows the chemical entity comparison view where the extracted chemical compound information from the researcher's uploaded document (left) is displayed side-by-side with one of the recommended papers (right). Molecular compounds are aligned and highlighted in green if they occur both in the researcher's input paper and in the recommended paper currently being explored. A deeper shade of green is used if a chemical entity match occurs with a higher frequency in the two papers. Compounds that occur only in the input paper or the recommended paper, and thus do not match are shown on a white background. A scrollbar lets researchers browse all extracted entities among the two papers. From a 'document selection' tab (top left), a researcher can choose a different paper from their collection of bookmarked recommendations to inspect in the chemical entities exploration view.

The ChemVis frontend is embedded in the RecVis frontend as shown in Figure 2. We introduced the RecVis research paper recommendation system in previous work [16]. The recommendation approach of RecVis accounts for a variety of paper features to compute paper similarity, such as academic citations, mathematical expressions, keywords, or figures. These other document features can also be seen in the gray bar at the top in Figure 3.

| | Input Paper | | | | Recommended Paper | | | |
|---|---|---|---|---|---|---|---|---|
| CID | Name | Structure | Molecular Formula | Molecular Weight | CID | Name | Structure | Molecular Formula | Molecular Weight |
| 10340 | Sodium carbonate | | Na₂CO₃ | 105.988 | 10340 | Sodium carbonate | | Na₂CO₃ | 105.988 |
| 24083 | Magnesium sulphate | | MgSO₄ | 120.37 | 24083 | Magnesium sulphate | | MgSO₄ | 120.37 |
| 962 | Water | | H₂O | 18.015 | 887 | Methanol | | CH₄O | 32.042 |

Figure 3: ChemVis user interface for chemical entity examination and comparison

Furthermore, users from different STEM domains can assign unique weights to similarity functions and thus customize the recommendations to their domain within RecVis.

By having introduced chemical entity recognition and visualization with ChemVis, we have made the first step towards expanding the capabilities of RecVis in order to better support researchers from the biomedical and chemistry domains.

Our code is available as open source from:
https://github.com/ag-gipp/chem_formula_extractor

**Future Work**

In the future, we plan to expand upon the functionality of *ChemVis* by also making use of ChemSchematicResolver and ReactionDataExtractor to extract additional chemical entity formats. Currently, if an extracted chemical entity is not found in the PubChem database, the entity is not visualized. To tackle this shortcoming, we are making the search process more robust and reliable. We also plan to extract and visualize any chemical properties mentioned in the full text, instead of retrieving this data only from the PubChem database.

Since chemical entities can be compared between different documents, we plan on testing different algorithms for recommending papers based on the contained chemical entities. These algorithms could range from simple matching to structural comparisons or matching of the chemical's metadata.

We would like to test different weightings of chemical entity similarity and textual similarity and assess the recommendation performance using chemists and biomedical researchers. In future work, the order or placement of chemical entities within the publications might also be considered when recommending chemistry publications. Another research avenue would be to use the chemical entity information extracted from research publications to improve expert recommender systems in the chemistry domain. The result could be a hybrid approach that considers all extracted chemical entity data in addition to traditional key phrase based similarity, as is standard for expert recommendation [17].

## 4 Conclusion

We presented the conceptualization and implementation of a paper recommender system tailored to chemists and biomedical researchers, which accounts for and visualizes chemical entities in their various formats. While searchable molecular entity databases exist, such as PubChem or ChemSpider, no paper recommendation system thus far accounts for and visualizes the chemical entity information contained in academic literature. We identified this research gap and used ChemDataExtractor v2, a toolkit for the reliable automated extraction of chemical data from research papers, to create a chemical-entity-aware paper recommendation interface. Our proposed prototype, *ChemVis* performs formula disambiguation and highlights semantically equivalent chemical entity information among recommended papers no matter whether a molecule or compound was formatted as a molecular formula, systematic name, generic name, or brand name. We showcased how an interactive visualization of chemical entities could support chemists in more quickly browsing the entity-relevant information in a side-by-side visualization. We hope our work inspires research on other specialized recommendation systems and interfaces tailored to the needs of researchers from a variety of domains.

## ACKNOWLEDGMENTS

This research was supported by the German Research Foundation (DFG) Grant No.: GI 1259/1.